\documentclass[aps,prl,twocolumn,twoside,floatfix,superscriptaddress,preprintnumbers]{revtex4}

\usepackage{graphicx}
\usepackage{color}
\usepackage{rotating}

\newcommand{\ba}{\begin{array}}
\newcommand{\ea}{\end{array}}

\newcommand{\beq}{\begin{equation}}
\newcommand{\eeq}{\end{equation}}
\newcommand{\bea}{\begin{eqnarray}}
\newcommand{\eea}{\end{eqnarray}}

\newcommand{\im}{{\rm Im}}

\newcommand{\cO}{{\cal O}}

\newcommand{\cL}{{\cal L}}

\newcommand{\cH}{{\cal H}}

\definecolor{red}{cmyk}{0,1,1,0.4}

\newcommand{\realV}{V}

\begin{document}

\title{EDMs vs.~CPV in $B_{s,d}$ mixing in two Higgs doublet models with MFV}

\author{Andrzej J.~Buras}
\affiliation{Physik-Department, Technische Universit\"at M\"unchen,
D-85748 Garching, Germany}
\affiliation{TUM Institute for Advanced Study, Technische Universit\"at M\"unchen,
\\Arcisstr.~21, D-80333 M\"unchen, Germany}

\author{Gino~Isidori}
\affiliation{TUM Institute for Advanced Study, Technische Universit\"at M\"unchen,
\\Arcisstr.~21, D-80333 M\"unchen, Germany}
\affiliation{INFN, Laboratori Nazionali di Frascati, Via E.~Fermi 40, I-00044 Frascati, Italy}

\author{Paride~Paradisi}
\affiliation{Physik-Department, Technische Universit\"at M\"unchen,
D-85748 Garching, Germany}

\begin{abstract}
We analyze the correlations between electric dipole moments (EDMs)
of the neutron and heavy atoms and CP violation in $B_{s,d}$
mixing in two Higgs doublet models respecting the Minimal Flavour
Violation hypothesis, with flavour-blind CP-violating (CPV) phases.
In particular, we consider the case of flavour-blind CPV phases from
i) the Yukawa interactions and  ii) the Higgs potential.
We show that in both cases the upper bounds on the above EDMs do not
forbid sizable non-standard CPV effects in $B_{s}$ mixing.
However, if a large CPV phase in $B_s$ mixing will be confirmed, this
will imply EDMs very close to their present experimental bounds,
within the reach of the next generation of experiments, as well as
${\rm BR}(B_{s,d}\to\mu^+\mu^-)$ typically largely enhanced over its
SM expectation. The two flavour-blind CPV mechanisms can be distinguished
through the correlation between $S_{\psi K_S}$ and $S_{\psi\phi}$ that is
strikingly different if only one of them is relevant. Which of these two
CPV mechanisms dominates depends on the precise values of $S_{\psi\phi}$
and $S_{\psi K_S}$, as well as on the CKM phase (as determined by tree-level
processes). Current data seems to show a mild preference for a {\it hybrid}
scenario where both these mechanisms are at work.
\end{abstract}

\maketitle

\section{1.~~Introduction}\label{sec:intro}

In the last few years, the two $B$ factories have established that flavor-changing
and CPV processes of $B_d$ mesons are well described by the Standard Model (SM) up
to an accuracy of $(10-20)\%$. This observation, together with the
good agreement between data and SM expectations in the kaon system, implies tight
constraints on flavor-changing phenomena beyond the SM and a potential problem for
a natural solution of the hierarchy problem, that calls for new physics (NP) not far
from the electroweak scale~\cite{Isidori:2010kg}.

An elegant way to solve this problem is provided by the Minimal Flavor Violation (MFV) hypothesis~\cite{D'Ambrosio:2002ex} (see also~\cite{Chivukula:1987py,Buras:2000dm}),
where flavor-changing transitions in the quark sector are entirely controlled by
the two quark Yukawa couplings. Despite apparently being quite restrictive, the MFV
hypothesis does not forbid sizable deviations from the SM in specific channels.
This is particularly true in models with two or more Higgs doublets, because of
the possibility to change the relative normalization of the two Yukawa couplings~\cite{D'Ambrosio:2002ex,Isidori:2001fv,Dedes:2002er,Buras:2002vd,Isidori:2006pk,Hurth:2008jc}.

A particularly interesting set-up is obtained introducing flavour-blind CPV phases compatible
with the MFV symmetry principle~\cite{Kagan:2009bn,Colangelo:2008qp,Mercolli:2009ns,Paradisi:2009ey,Ellis:2007kb}.

As recently shown in~\cite{Buras:2010mh}, the general formulation of the MFV hypothesis with
flavour-blind CPV phases applied to two Higgs doublet models (2HDMs) is very effective in
suppressing flavour-changing neutral-currents (FCNCs) to a level consistent with experiments,
leaving open the possibility of sizable non-standard effects also in CPV observables.
In what follows, we will call this framework ${\rm 2HDM_{\overline{MFV}}}$ with 
the ``bar'' indicating flavour-blind CPV phases.

As discussed in~\cite{Buras:2010mh}, the ${\rm 2HDM_{\overline{MFV}}}$
can accommodate a large CPV phase in $B_s$ mixing, as hinted by CDF and
D0 data~\cite{Aaltonen:2007he,Abazov:2010hv,Abazov:2008fj}, while ameliorating
simultaneously the observed anomaly in the relation between $\epsilon_K$ and 
$S_{\psi K_S}$~\cite{Lunghi:2008aa,Buras:2008nn}.

On general grounds, it is natural to expect that flavour-blind CP phases contribute
also to CPV flavour-conserving processes, such as the EDMs. Indeed, the choice
adopted in~\cite{D'Ambrosio:2002ex} to assume the Yukawa couplings as the unique
breaking terms of both the flavour symmetry and the CP symmetry, was motivated
by possibly too large effects in EDMs with generic flavour-blind CPV phases.
This potential problem has indeed been confirmed by the recent model-independent
analysis in~\cite{Batell:2010qw}.

In this Letter we address the role of EDMs, and their correlation with CPV effects
in $B_{s}$ mixing, in the ${\rm 2HDM_{\overline{MFV}}}$.
Following the recent analysis in~\cite{Buras:2010mh}, we focus on the contributions
to these observables generated by the integration of the Higgs fields only, assuming
a high suppression scale for effective operators not induced by the Higgs exchange.

We analyse in particular two sources of CPV phases: flavour-blind phases in
i)~the Yukawa interactions and in ii)~the Higgs potential. Flavour-blind phases
in~the Yukawa interactions, together with the assumption of small $SU(2)_L$
breaking in the heavy Higgs sector, can lead to large corrections to $S_{\psi\phi}$.
In this case the new CPV effects in $B_{d}$ mixing are suppressed by a factor of 
$m_d/m_s$, compared to $B_{s}$ mixing, and go in the right direction to ameliorate
the prediction of $S_{\psi K_S}$~\cite{Buras:2010mh}. Flavour-blind phases of the
type~ii) can also affect $S_{\psi\phi}$, provided the $SU(2)_L$ breaking in the
heavy Higgs sector is not negligible. However, in this case the correction in $B_{s}$
and $B_d$ mixing is universal and the magnitude of the effect in $S_{\psi\phi}$
is bounded by the limited amount of NP allowed in~$S_{\psi K_S}$~\cite{Buras:2008nn,Kagan:2009bn,Blum:2010mj}.
Which of the two flavour-blind CPV mechanisms dominates depends on the value of
$S_{\psi\phi}$, which is still affected by a sizable experimental error, and also
by the precise amount of NP allowed in $S_{\psi K_S}$.

We show that in both cases the upper bounds on the neutron and heavy atom EDMs
do not forbid sizable non-standard CPV effects in $B_{s}$ mixing. Interestingly
enough, in both cases sizable CPV effects in $B_{s}$ mixing imply lower bounds
for these EDMs within the future experimental resolutions. Moreover, in both
cases ${\rm BR}(B_{s}\to\mu^+\mu^-)$ and ${\rm BR}(B_{d}\to\mu^+\mu^-)$ are
typically largely enhanced over their SM expectations.

Our paper is organized as follows. In Section 2 we recall briefly the most important
ingredients of the ${\rm 2HDM_{\overline{MFV}}}$, concentrating in particular
on the two new sources of CPV in question. In Sections 3,4, we exploit the sensitivity
of the EDMs and $B_{s,d}-\bar B_{s,d}$ mixing, respectively, to these new CPV phases.
The numerical analysis of various correlations that have been advertised in the abstract
is performed in Section 5. Here we demonstrate that the current data seems to show a
mild preference for a {\it hybrid} scenario where both new mechanisms of CPV are at work.

\boldmath
\section{2.~~The ${\rm 2HDM_{\overline{MFV}}}$}
\unboldmath

In the following, we consider a 2HDM supplemented by the MFV hypothesis, where the Yukawa
matrices are the only sources of breaking of the $SU(3)_q$ flavour group, but they are not
the only allowed sources of CP violation~\cite{Kagan:2009bn,Mercolli:2009ns,Paradisi:2009ey}.

The most general renormalizable and gauge-invariant Yukawa interactions in a 2HDM are
\bea
- \cL_Y^{\rm gen} &=& 
\bar Q_L X_{d1} D_R H_1 + \bar Q_L X_{u1} U_R H_1^c
\nonumber \\
&+& \bar Q_L X_{d2} D_R H_2^c + \bar Q_L X_{u2} U_R H_2 +{\rm h.c.}~,
\label{eq:generalcouplings}
\eea
where the Higgs fields $H_{1,2}$ have hypercharges $Y=\pm 1/2$, $H_{1(2)}^c=-i\tau_2 H_{1(2)}^*$
and the $X_i$ are $3\times 3$ matrices with a generic flavour structure. We also assume real vevs
for the two fields, $\langle H^\dagger_{1(2)} H_{1(2)} \rangle=v^2_{1(2)}/2$, with $v^2 =v_1^2+v_2^2 \approx (246~{\rm GeV})^2$ and, for later purpose, we define $c_\beta =v_1/v$, $s_\beta =v_2/v$,
$t_\beta = s_\beta/c_\beta$.

The general structure implied by the MFV hypothesis for the renormalizable Yukawa
couplings $X_{di}$ and $X_{ui}$ is a polynomial expansions in terms of the two
(left-handed) spurions $Y_u Y_u^\dagger$ and $Y_d Y_d^\dagger$~\cite{D'Ambrosio:2002ex,Buras:2010mh}:
\bea
X_{d1} &=& Y_d~, \nonumber \\
X_{d2} &=& \epsilon_{0} Y_d + \epsilon_{1} Y_d  Y_d^\dagger Y_d
        +  \epsilon_{2} Y_u Y_u^\dagger Y_d + \nonumber \\
       &&+ \epsilon_{3} Y_u Y_u^\dagger Y_d  Y_d^\dagger Y_d
         +  \epsilon_{4} Y_d  Y_d^\dagger Y_u Y_u^\dagger Y_d + \ldots~,
\nonumber \\
X_{u2} &=& Y_u~, \nonumber \\
X_{u1} &=& \epsilon^\prime_{0} Y_u + \epsilon^\prime_{1}  Y_u Y_u^\dagger Y_u
+  \epsilon^\prime_{2}  Y_d Y_d^\dagger Y_u + \ldots~,
\label{eq:XMFVgen}
\eea
where the $\epsilon^{(\prime)}_{i}$ are complex parameters. 
We work under the assumption $\epsilon^{(\prime)}_{i}\ll 1$,
as expected by an approximate $U(1)_{\rm PQ}$ symmetry
that forbids non-vanishing $X_{u1}$ and $X_{d2}$ at the tree level.
We also assume negligible violations of the $U(1)_{\rm PQ}$
symmetry in the lepton Yukawa couplings. 

After diagonalising quark mass terms and rotating the Higgs fields such that only one
doublet has a non-vanishing vev, the interaction of down-type quarks with the neutral
Higgs fields assumes the form
\beq
\cL_{\rm n.c.}^{d} = -  \frac{\sqrt{2}}{v} \bar d_L M_d d_R \phi_v^0
- \frac{1}{s_\beta}{\bar d}_L Z^{d}
\lambda_d d_R \phi_H^0 {\rm +h.c.}\,,
\label{eq:LH_FCNC}
\eeq
where $\phi_v$ ($\phi_H$) is the Higgs doublet with non-vanishing (vanishing)
vev $\langle\phi^0_v\rangle =v/\sqrt{2}$ ($\langle\phi^0_H\rangle =0$) and
$\phi_H^0 = (H+iA)/\sqrt{2}$ with $H$ ($A$) being the CP-even (CP-odd) heavy
Higges, if the Higgs potential is CP-invariant. The flavour structure of the
$Z^d$ couplings, which play a key role in our analysis, is
\beq
Z^d_{ij} =
\bar{a}\delta_{ij} +
\left[
a_0 V^\dagger \lambda_u^2 V + a_1 V^\dagger \lambda_u^2 V
\Delta + a_2 \Delta V^\dagger \lambda_u^2 V
\right]_{ij}\,,
\label{eq:C^S0}
\eeq
where $V$ is the physical CKM matrix, $\Delta \equiv \rm{diag}(0,0,1)$ and $\lambda_{u,d}$
are the diagonal up Yukawa couplings in the limit $\epsilon^{(\prime)}_i \to 0$ 
(see~\cite{D'Ambrosio:2002ex,Buras:2010mh} for notations).
As explicitly given in~\cite{D'Ambrosio:2002ex,Buras:2010mh}, the $a_i$ are flavour-blind
coefficients depending on the $\epsilon_i$, on $t_\beta$, and on the overall normalization
of the Yukawa couplings. Even if  $\epsilon^{(\prime)}_{i}\ll 1$,
the $a_i$ can reach values of $\cO(1)$ at large $t_\beta$ and can
be complex if we allow flavour-blind phases in the model.

Similarly, the charged-current interactions of the fermions with the charged Higgs are
parameterized by the following flavour changing effective Lagrangian~\cite{D'Ambrosio:2002ex}
\beq
\cL_{\rm H^+ } =
\frac{1}{c_{\beta}}
\Big[ {\bar U}_L C_{R}^{H^+} \lambda_d D_R +
      \frac{1}{t^{2}_{\beta}} {\bar U}_R  \lambda_u C_{L}^{H^+} D_L
\Big] H^+ {\rm +h.c.}~,
\label{eq:LH_charged}
\eeq
where the flavour structure of the $C_{R,L}^{H^+}$ is
\bea
C_{R}^{H^+} &=& \left( b_0 \realV + b_1 \realV \Delta + b_2 \Delta\realV + b_3 \Delta\right)~,
\label{eq:C_R^H+}
\\
C_{L}^{H^+} &=& \left( b'_0 \realV + b'_1 \realV \Delta + b'_2 \Delta \realV + b'_3 \Delta \right)\,.
\label{eq:C_L^H+}
\eea
In analogy to the  $a_i$, also the $b_i$ and $b^{'}_{i}$ coefficients are flavour-blind, 
naturally of $\cO(1)$, and possibly complex.

Another source of CP violation in the ${\rm 2HDM_{\overline{MFV}}}$, that is relevant
for our analysis, arises from the Higgs potential~\cite{Weinberg:1989dx,Barr:1990vd,Barr}
(see also~\cite{Gorbahn:2009pp}). We recall that the most general 2HDM potential that is renormalizable and gauge invariant is~\cite{Gunion:1989we}
\begin{eqnarray}
&& \!\!\!\!\!\!\!\!\!
V = \mu_1^2|H_1|^2 + \mu_2^2|H_2|^2 + (b H_1 H_2+{\rm h.c})+
\nonumber\\
&& \!\!\!\!\!\!
+\frac{\lambda_1}{2}|H_1|^4 
+\frac{\lambda_2}{2}|H_2|^4+\lambda_3|H_1|^2|H_2|^2 + \lambda_4|H_1H_2|^2 +
\nonumber\\
&& \!\!\!\!\!\!
+\! \left[\frac{\lambda_5}{2}(H_1H_2)^2 + \lambda_6|H_1|^2H_1H_2 + \lambda_7|H_2|^2H_1H_2 +
{\rm h.c}\right]\,,
\nonumber
\end{eqnarray}
where $H_1 H_2= H_1^T (i \sigma_2) H_2$. All the parameters must be real with the exception of
$b$ and $\lambda_{5,6,7}$. Exploiting the freedom to change the relative phase between $H_1$
and $H_2$, we can cancel the phase of $b$ and $\lambda_{6,7}$ relative to $\lambda_{5}$.
Moreover, the coefficients $\lambda_{6,7}$ can be set to zero imposing a discrete $Z_2$ 
symmetry that is only softly broken by the terms proportional to $b$ and $\lambda_5$.

In order to simplify the discussion, without loosing generality as far as the CP properties
are concerned, we set $\lambda_3=\lambda_4=\lambda_6=\lambda_7=0$ and choose the basis where
only $\lambda_5$ is complex. The resulting spectrum contains a charged Higgs, with the mass
\beq
M_{H^\pm}^2=\frac{b}{c_\beta s_\beta}-\frac{\text{Re}(\lambda_5)}{2}v^2~,
\eeq
and three neutral Higgses with masses
\bea
M_{1}^{2}&\simeq&\lambda_2 v^2
\nonumber\,,\\
M_{2(3)}^{2}&\simeq&\frac{b}{c_\beta}-\frac{v^2}{2}\left(\text{Re}(\lambda_5)
\mp |\lambda_5|\right)~,
\label{eq:mass_S0}
\eea
where in Eqs.~(\ref{eq:mass_S0}) we have assumed $t_\beta\gg 1$. Notice the approximate
degeneracy of the charged Higgs and the two neutrals of mass $M_{2}$ and $M_{3}$ in the
limit $\lambda_5\to 0$.

In the absence of CP violation, the physical Higgs eigenstates are given by the two CP-even
fields $h,H$ and by the CP-odd field $A$. In the presence of CP violation, $h,H,A$ are mixed
and the mass eigenstates are not anymore CP eigenstates. Still, it is convenient to write
the Higgs potential in terms of the fields $h,H,A$. It turns out that
\bea
 V &=& \frac{M^2_h}{2} h^2 + \frac{M^2_{H}}{2} H^2 +\frac{M^2_A}{2} A^2
+ M^2_{H^\pm} {H^+ H^-}+
\nonumber\\
&+& \langle Ah\rangle\, Ah + \langle AH\rangle\, AH + ....\,,
\eea
where $\langle Ah\rangle$ and $\langle AH\rangle$ read
\begin{eqnarray}
\label{eq:AH_mixing_2hdm}
\langle Ah\rangle &=& - \frac{v^2}{2} c_{\beta}\, \text{Im} { \lambda_5}\,,
\\
\langle AH\rangle &=& - \frac{v^2}{2} s_{\beta}\, \text{Im} { \lambda_5}\,.
\end{eqnarray}
Notice that in the large $t_\beta$ regime the mixing $\langle Ah\rangle$ is negligible,
in contrast to $\langle AH\rangle$. Moreover, if $\langle AH\rangle \ll M^{2}_{A,H}$,
as it happens in the so-called decoupling regime, we can still treat the fields $h,H,A$
as approximate mass-eigenstates and the mixing $\langle AH\rangle$ can be parameterized
as an effective mass insertion in the scalar propagator.

\section{3.~~Electric Dipole Moments}

Among the various atomic and hadronic EDMs, the Thallium, neutron and Mercury EDMs represent
the most sensitive probes of CP violating effects (see table~\ref{tab:observables}).
\begin{table*}
\centering
\begin{tabular}{|l|l|l|l|}
\hline
Observable & Exp. Current & Exp. Future \\
\hline\hline
$|d_{Tl}|~~[e\,$cm] & $< 9.0 \times 10^{-25}$~\cite{Regan:2002ta} & ~$\approx 10^{-29}$~\cite{Pospelov:2005pr} \\
\hline
$|d_{Hg}|~~[e\,$cm] & $< 3.1 \times 10^{-29}$~\cite{Griffith:2009zz} & ~?~ \\
\hline
$|d_n|~~[e\,$cm] & $< 2.9 \times 10^{-26}$~\cite{Baker:2006ts} & ~$\approx 10^{-28}$~\cite{Pospelov:2005pr} \\
\hline
\end{tabular}
\caption{Current/expected experimental sensitivities for the most relevant EDMs.}
\label{tab:observables}
\end{table*}
The effective CP-odd Lagrangian describing the quark (C)EDMs, that is relevant for
our analysis, reads
\bea
-{\cal L}_{\rm eff} &=&
\sum_{f} i \frac{d_f}{2}\bar{f} (F\sigma)\gamma_5 f
+ \sum_{f} i \frac{d^c_f}{2} g_s \bar{f} (G\sigma)\gamma_5 f
\nonumber\\
&+&
\sum_{f,f'}C_{ff'}(\bar{f}f)(\bar{f'}i\gamma_5f')\,,
\label{Eq:CPodd}
\eea
where $F_{\mu\nu}$ ($G^a_{\mu\nu}$) is the electromagnetic (chromomagnetic) field strength,
$d^{(c)}_{f}$ stands for the quark (C)EDMs while $C_{ff'}$ is the coefficient of the CP-odd
four fermion interactions.

The thallium EDM ($d_{\rm{Tl}}$) can be estimated as~\cite{KL,Pospelov:2005pr,Demir:2003js}
\begin{equation}
d_{\rm Tl} \simeq -585\cdot d_e - e\,\left(43\rm{GeV}\right)\,C_{S}\,,
\end{equation}
where $d_e$ is the electron EDM while $C_S$ stems from the CP-odd four fermion interactions
and reads~\cite{Demir:2003js}
\bea
C_S &\simeq&
C_{de}\frac{29\,{\rm MeV}}{m_d} +
C_{se}\frac{k \times 220\,{\rm MeV}}{m_s} +
\nonumber\\
 &+& C_{be}\frac{ 66\,{\rm MeV}(1-0.25 k)}{m_b}\,,
\label{eq:CS}
\eea
with $\kappa \simeq 0.5 \pm 0.25$ \cite{KBM}.

\begin{figure}[t]
\centering
\includegraphics[scale=1.0]{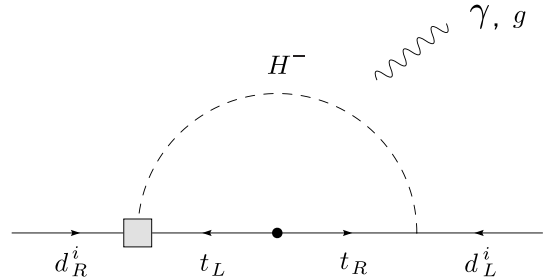}
\caption{\label{fig2} Charged-Higgs mediated diagrams
contributing to the down-quark (C)EDMs.}
\end{figure}

The neutron EDM $d_n$ can be estimated from the naive quark model as
$d_n\approx\frac{4}{3} d_d-\frac{1}{3} d_u$, where $d^{(c)}_{f}$ are evaluated
at $1$~GeV by means of QCD renormalization~\cite{Degrassi:2005zd},
starting from the corresponding values at the electroweak scale.
The alternative estimate we use in our numerical analysis is the one 
obtained from QCD sum rules~\cite{qcdsumrules,Demir:2002gg,Demir:2003js,Olive:2005ru},
which leads to
\beq
 d_n \!= (1\pm 0.5)\Big[ 1.4 (d_d-0.25 d_u) + 1.1 e\, (d^c_d+0.5 d^c_u)\Big]
\label{Eq:dn_odd}
\eeq
and
\bea
\label{eq:mercuryedm}
d_{\rm Hg}
&\simeq&
7\times 10^{-3}\,e\,(d_u^c-d_d^c) + 10^{-2}\,d_e +
\nonumber\\
&+& e\,\left( 3.5 \times 10^{-3}\,\rm{GeV} \right)\,C_{S} +
\nonumber\\
&-& e\left( 1.4 \times 10^{-5}\rm{GeV}^2 \right)\times
\nonumber\\
&\times&
\!\!\!\left[0.5\frac{C_{dd}}{m_d} + 3.3k\frac{C_{sd}}{m_s} + (1-0.25 k)\frac{C_{bd}}{m_b}\right]
\eea
for the Mercury EDM~\cite{Demir:2003js,Olive:2005ru}. In these numerical formulae the
$d^{(c)}_{f}$ are evaluated at $1$~GeV. In the following, we provide the expressions for
$d^{(c)}_{f}$ and $C_{ff'}$ at the electroweak scale in the context of a 2HDM, starting
from the general formulae of Ref.~\cite{Ellis:2008zy,Hisano:2006mj} obtained in the context
of Supersymmetry.

The CP violating effects arising from CP-odd CP-even scalar mixing and their impact on
the EDMs were studied previously in the context of 2HDMs with spontaneous breaking of
CP in refs.~\cite{Weinberg:1989dx,Barr,Barr:1990vd}.

For $C_{ff'}$ one has
\beq
\label{eq:cff}
C_{ff'} = \frac{m_{f}\,m_{f'}}{v^2}\,
\frac{{\rm Im}\,\omega_{ff'}}{M^2_A}\,t^{2}_{\beta}\,,
\eeq
where we have defined
\beq
\label{eq:CPV_source}
{\rm Im}\,\omega_{ff'}
\simeq
{\rm Im}\left(Z^{d\star}_{ff}\,Z^{d}_{f'f'}\right) +
\frac{\langle AH\rangle}{M_{A}^{2}}\,{\rm Re}\left(Z^{d}_{ff}\,Z^{d}_{f'f'}\right)\,.
\eeq
Notice that in Eq.~(\ref{eq:CPV_source}) we have assumed that
$M_{H}^{2}\simeq M_{A}^{2}\gg\langle AH\rangle$. The explicit
expressions for the $\omega_{ff'}$ relevant for our analysis are
\bea
{\rm Im}\,\omega_{ed} &=& {\rm Im}\,\omega_{es} =
{\rm Im}\,\sigma + \frac{\langle AH\rangle}{M_{A}^{2}}\,{\rm Re}\,\sigma\,,
\label{eq:downtypeomegai}
\\
{\rm Im}\,\omega_{eb}&=&
{\rm Im}\,\xi + \frac{\langle AH\rangle}{M_{A}^{2}}\,{\rm Re}\,\xi\,,
\\
{\rm Im}\,\omega_{dd}&=&{\rm Im}\,\omega_{ds}=
\frac{\langle AH\rangle}{M_{A}^{2}}\,{\rm Re}\,(\sigma^2)\,,
\\
{\rm Im}\,\omega_{db}&=& 
{\rm Im}\,(\sigma^{\star}\xi)+\frac{\langle AH\rangle}{M_{A}^{2}}\,
{\rm Re}\,(\sigma\xi)\,,
\label{eq:downtypeomegaf}
\eea
where we have defined
\bea
\xi    &=& \overline{a} + (a_0 + a_1 + a_2)\lambda_t^2 \,,\\
\sigma &=& \overline{a} + |V_{tq}|^2 \lambda_t^2 a_0\qquad\qquad (q=d,s)~.
\eea
Concerning the quark (C)EDMs, they are induced already at the one-loop level by means
of the exchange of the charged-Higgs boson and top quark~\cite{Hisano:2006mj}, as it
is shown in Fig.~\ref{fig2}.

In our scenario, their explicit expressions read
\bea
\left\{\frac{d_{d}}{e},~d_{d}^c\right\}
&=&
\frac{\alpha_2}{4\pi}\frac{m_{d}}{m_W^2} \frac{m^2_t}{M^{2}_{H^\pm}}
|V_{td}|^2 F_{7,8}(x_{tH})\times
\nonumber\\
&\times&\im\left[\,(b_{0}^{*}+b_{2}^{*})(b_{0}^{\prime}+b_{2}^{\prime})\,\right]\,,
\eea
where $x_{tH}=m^2_t/M^{2}_{H^\pm}$ and the expressions for the loop functions
$F_{7,8}$ are given in~\cite{Hisano:2006mj}.

Notice that the above effects are CKM suppressed by the factor $|V_{td}|^2\approx 10^{-4}$.
In particular, even in the most favourable case where $m_t=M_{H^\pm}$ and
$\im\left[(b_{0}^{*}+b_{2}^{*})(b_{0}^{\prime}+b_{2}^{\prime})\right]= \mathcal{O}(1)$,
it turns out that $|d_{d}/e|\approx|d^{c}_{d}|= \mathcal{O}(10^{-27})$, leading to
predictions for $d_n$ and $d_{\rm Hg}$ well under control (though observable with
improved experimental resolutions). Therefore, since the $b_i$'s do not enter
the predictions for $B_{s,d}$ mixings, we neglect the above one-loop effects to
$d_n$ and $d_{\rm Hg}$ in our numerical analysis.

As a result, two loop contributions might compete or even dominate over one loop-effects
provided they overcome the strong CKM suppression.

Indeed, this is the case for the two-loop Barr--Zee contributions~\cite{Barr:1990vd}
to the fermionic (C)EDMs (see Fig.~\ref{barr_zee}) that read
\bea
\frac{d_f}{e} \!\!\!&=&\!\!\!
-\sum_{q=t,b}\!\!
\frac{N_c q_f \alpha_{\rm em}^2\,q_q^2\,m_f}{8\pi^2s_W^2M_W^2}
\frac{m_q^2}{M_{A}^2}\,t^{0,2}_{\beta}\times
\nonumber\\
&\times&
\left[
f(\tau_{q})\,{\rm Im}\,\omega_{qf} + g(\tau_{q})\,{\rm Im}\,\omega_{fq}
\right]\,,
\eea
\bea
d_{f}^c \!\!\!&=&\!\!\!
-\sum_{q=t,b}
\frac{\alpha_s\,\alpha_{\rm em}\,m_{f}}{16\pi^2s_W^2M_W^2}
\frac{m_q^2}{M_{A}^2}\,t^{0,2}_{\beta}\times
\nonumber\\
&\times&
\left[
f(\tau_{q})\,{\rm Im}\,\omega_{qf} + g(\tau_{q})\,{\rm Im}\,\omega_{fq}
\right]
\,,
\eea
where $q_{\ell}$ is the electric charge of the fermion $\ell$, $\tau_{q}=m_q^2/M_{A}^2$
and $f(\tau)$, $g(\tau)$ are the two-loop Barr--Zee functions defined
in~\cite{Barr:1990vd,Ellis:2008zy}. In analogy to the down-type $\omega_{ff'}$
in Eqs.~(\ref{eq:downtypeomegai})--(\ref{eq:downtypeomegaf}), we have defined
\bea
{\rm Im}\,\omega_{et}&=& 
\frac{\langle AH\rangle}{M_{A}^{2}}\,,
\\
{\rm Im}\,\omega_{dt}&=&
{\rm Im}\,\sigma^{\star} + \frac{\langle AH\rangle}{M_{A}^{2}}\,{\rm Re}\,\sigma\,.
\eea
\begin{figure}[t]
\centering
\includegraphics[scale=0.6]{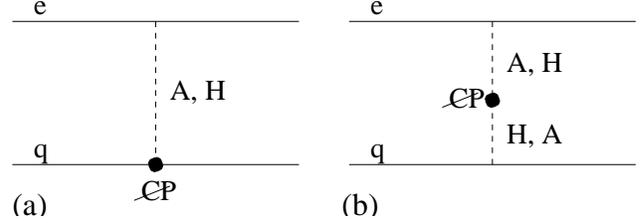}
\caption{\label{fig3} Examples of Higgs-mediated four-fermion interactions with
CP violation in the Higgs-fermion vertex (a) and on the Higgs propagator (b).}
\end{figure}

\begin{figure}[t]
\centering
\includegraphics[scale=0.8]{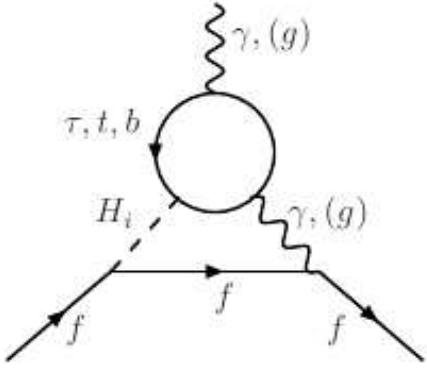}
\caption{\label{barr_zee} Example of Barr-Zee diagrams contributing to the
fermion (C)EDM in a 2HDM from neutral Higges $H_i$. CP violation is assumed
in the Higgs-fermion vertex and on the Higgs propagator as in Fig.\ref{fig3}.}
\end{figure}
%

%
\begin{figure}
\centering
\includegraphics[scale=0.4]{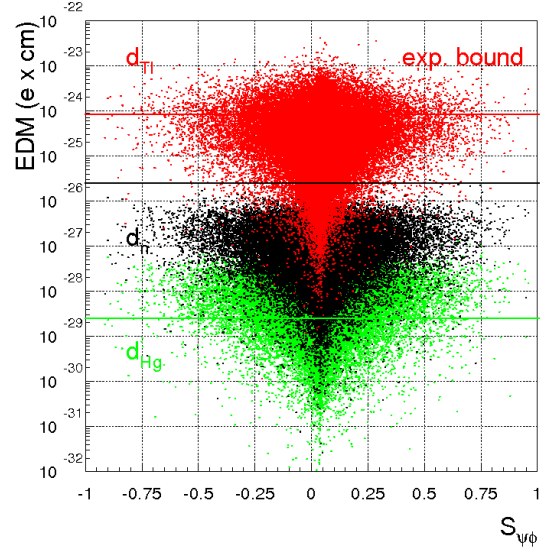}
\includegraphics[scale=0.4]{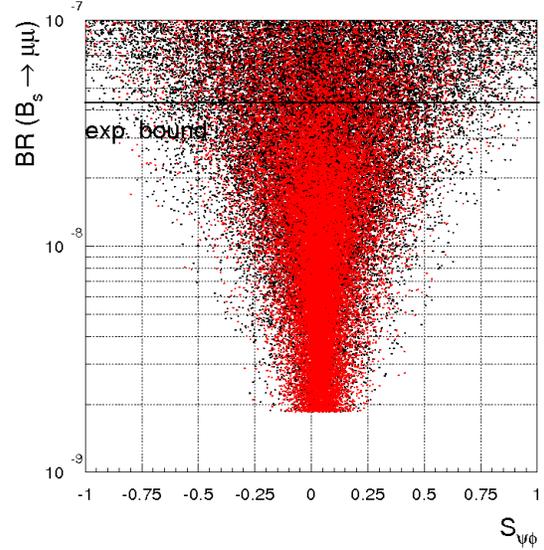}
\caption{
Upper: EDMs vs.~$S_{\psi\phi}$. Lower: $B_s\to\mu^+\mu^-$ vs.~$S_{\psi\phi}$.
Red dots fulfill the EDM constraints while the black ones do not.
In both plots we have assumed $(|\overline{a}|,|a_0|,|a_1|,|a_2|)< 2$,
$\lambda_5=0$, and $0<(\phi_{\overline{a}},\phi_{a_0},\phi_{a_1},\phi_{a_2})< 2\pi$.}
\label{edm_1}
\end{figure}
\begin{figure}
\centering
\includegraphics[scale=0.4]{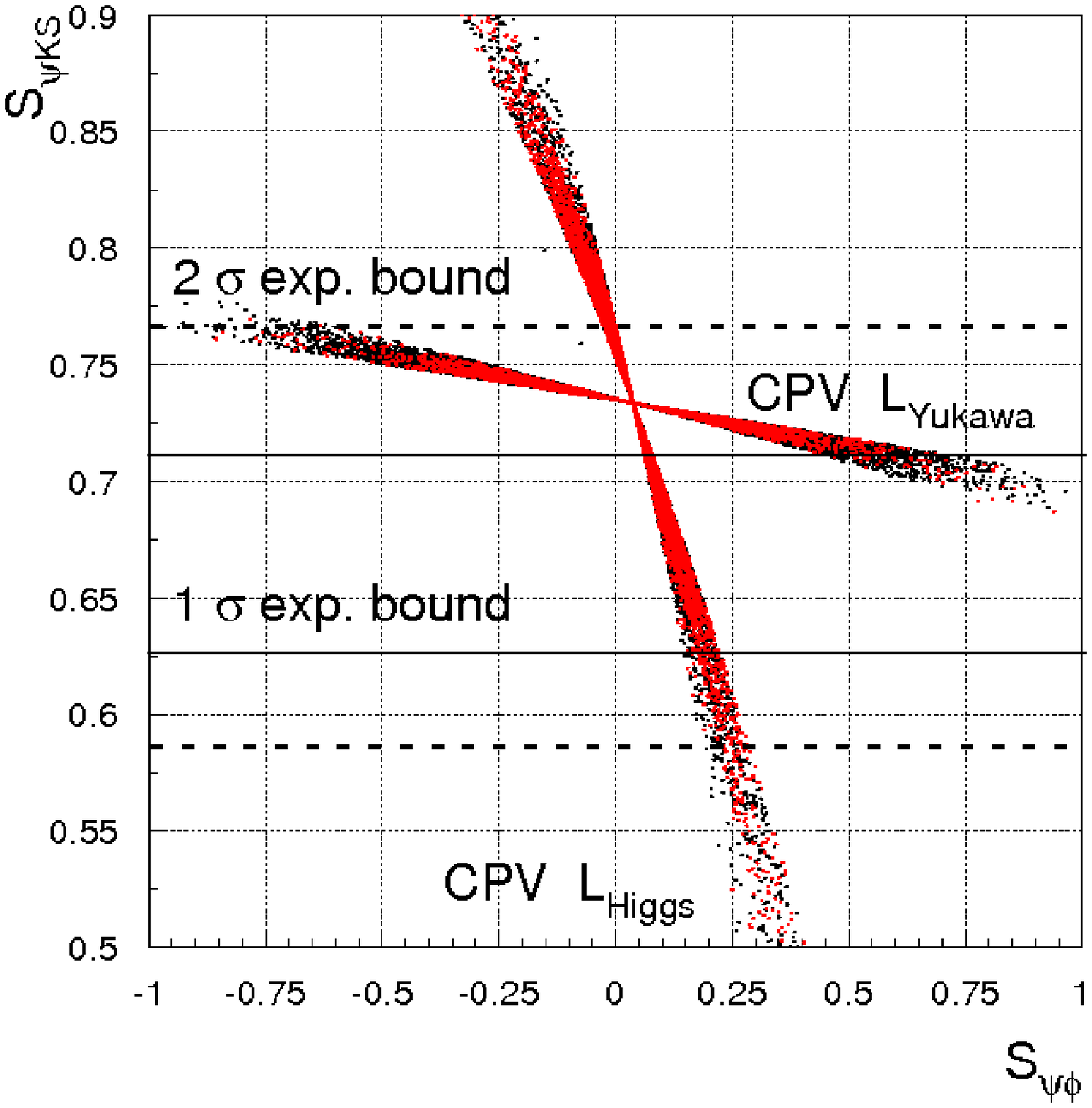}
\includegraphics[scale=0.4]{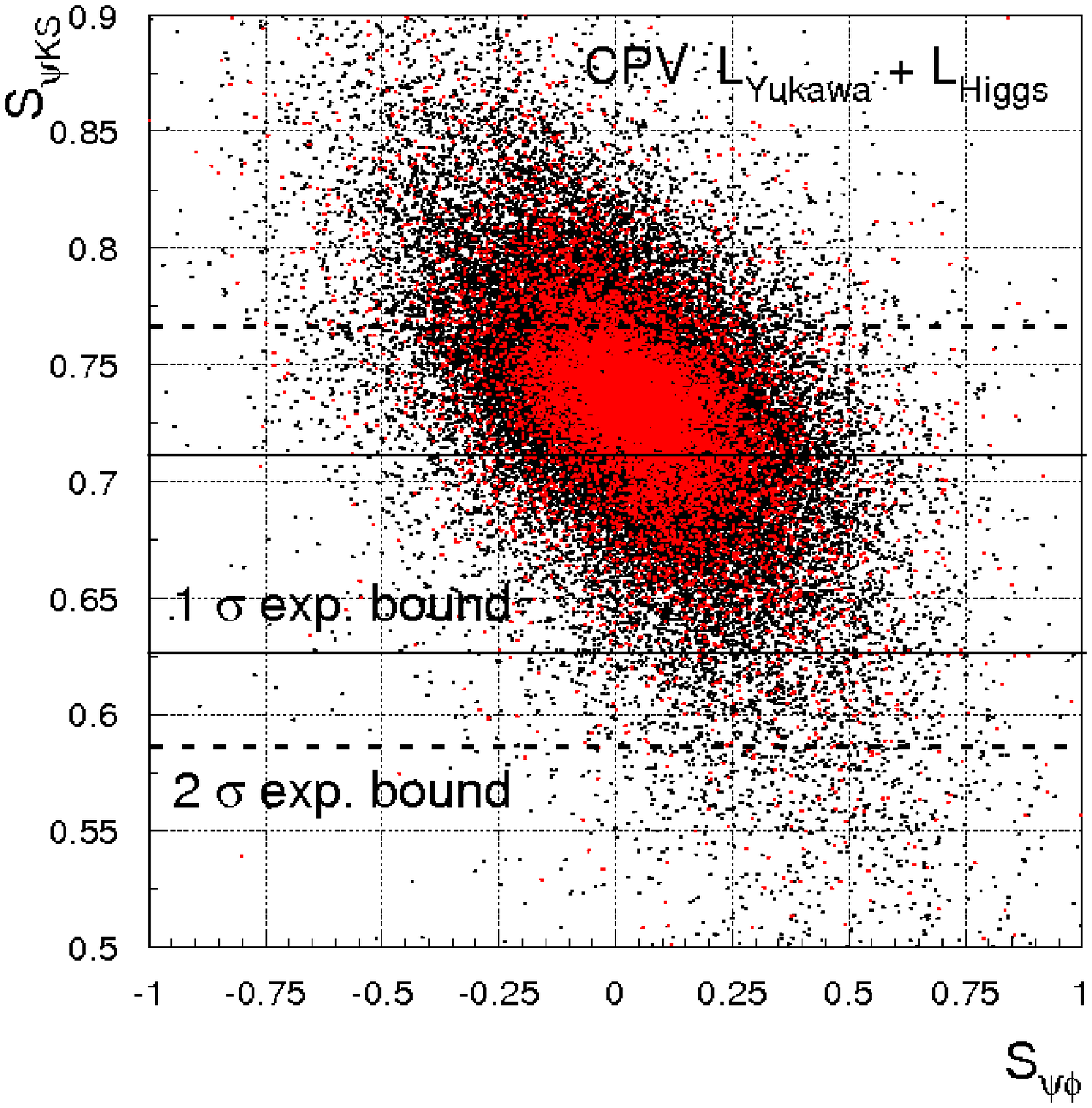}
\caption{
Correlation between $S_{\psi\phi}$ and $S_{\psi K_S}$ with CPV sources in the Yukawa
couplings and Higgs potential switched on separately (upper plot) or simultaneously
(lower plot). In all plots we have assumed $(|\overline{a}|,|a_0|,|a_1|,|a_2|)< 2$.
In the upper plot, the ``CPV $\cL_{\rm Yukawa}$'' points are obtained for $\lambda_5=0$
and $0<(\phi_{\overline{a}},\phi_{a_0},\phi_{a_1},\phi_{a_2})<2\pi$, while the
``CPV $\cL_{\rm Higgs}$'' points are obtained for
$(\phi_{\overline{a}},\phi_{a_0},\phi_{a_1},\phi_{a_2})=0$, $|\lambda_5|=0.1$
and $0<\phi_{\lambda_5}<2\pi$. In the lower plot $0<(\phi_{\overline{a}},\phi_{a_0},\phi_{a_1},\phi_{a_2})<2\pi$, $|\lambda_5|=0.1$
and $0<\phi_{\lambda_5}<2\pi$. Red dots fulfill the EDM constraints while the black
ones do not.}
\label{edm_2}
\end{figure}
%

%
\begin{figure}
\centering
\includegraphics[scale=0.4]{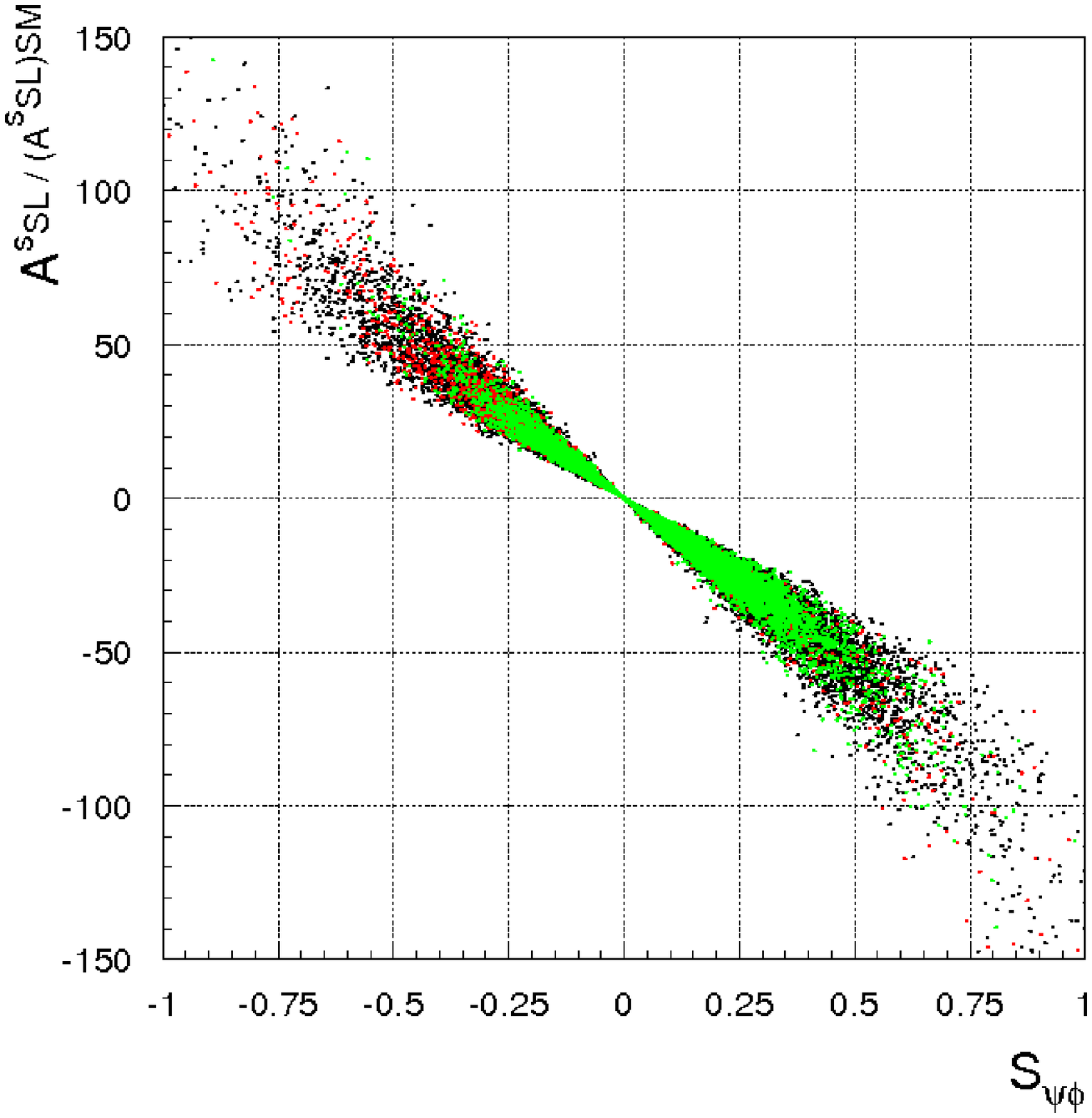}
\caption{
Correlation between $A^{s}_{SL}/A^{s}_{SL}({\rm SM})$ and $S_{\psi\phi}$
where $A^{s}_{\text{SL}}({\rm SM})=(2.6 \pm 0.5) \times 10^{-5}$~\cite{Ciuchini:2003ww}.
In the plot we have assumed $(|\overline{a}|,|a_0|,|a_1|,|a_2|)< 2$, $0<(\phi_{\overline{a}},\phi_{a_0},\phi_{a_1},\phi_{a_2})<2\pi$, $|\lambda_5|=0.1$
and $0<\phi_{\lambda_5}<2\pi$. Red dots fulfill the EDM constraints while the
black ones do not. Green dots further satisfy the constraint from $S_{\psi K_S}$
at the $95\%$ CL..}
\label{edm_3}
\end{figure}

To have an idea of where we stand, let us evaluate separately the contributions to the
physical EDMs arising from the fermionic (C)EDMs and from the four-fermion operator.

Assuming the example where $m_A=500$~GeV and $t_{\beta}=10$, the electron EDM contribution
to $d_{\rm Tl}$ reads
\bea
-\frac{d_{\rm Tl}(d_e)}{e~{\rm cm}}\approx
2\!\times\!10^{-26}\,\text{Im}\,\xi
+2\!\times\!10^{-24}\,\frac{\langle AH\rangle}{M_{A}^{2}}\,,
\eea
while $d^c_d$ generates the following $d_{\rm Hg}$
\bea
\frac{d_{\rm Hg}(d^c_d)}{e~{\rm cm}}
\!\approx\!
4\!\times\! 10^{-29}\,\text{Im}\,(\sigma^{\star}\xi)+
10^{-27}\,
\frac{\langle AH\rangle}{M_{A}^{2}}\,\text{Re}\,\sigma\,.
\eea
The down quark (C)EDMs generate the following $d_n$
\bea
-\frac{d_{n}(d^{(c)}_{d})}{e~{\rm cm}}
\!\approx\!
10^{-27}\text{Im}\,(\sigma^{\star}\xi)\!+\!
2\!\times\!10^{-26}
\frac{\langle AH\rangle}{M_{A}^{2}}\text{Re}\,\sigma\,.
\eea
Finally, let us consider $d_{\rm Tl}$ as induced by the four-fermion effects via $C_S$
\bea
\frac{d_{\rm Tl}(C_S)}{e~{\rm cm}}
\!\!&\approx&\!\!
10^{-25}\,(1+7.6\,k)
\left[\text{Im}\,\sigma - \frac{\langle AH\rangle}{M_{A}^{2}}\,\text{Re}\,\sigma\right]+
\nonumber\\
&+&
2.3\!\times\!10^{-25}\,
\left[\text{Im}\,\xi - \frac{\langle AH\rangle}{M_{A}^{2}}\,\text{Re}\,\xi\right]\,,
\eea
and similarly $d_{\rm Hg}(C_S)\approx -10^{-4}\!\times\!d_{\rm Tl}(C_S)$.

\section{4.~~$\Delta F=2$ amplitudes}

The low-energy operators that are relevant for $\Delta B=2$ transitions are~\cite{Buras:2001ra}
\bea
Q_1^{VLL}&=&(\bar b_L \gamma_\mu q_L)(\bar b_L \gamma^\mu q_L)\,, \\
Q_1^{SLL}&=&(\bar b_R q_L)(\bar b_R q_L)\,, \\
Q_2^{SLL}&=&(\bar b_R \sigma_{\mu\nu} q_L)(\bar b_R \sigma^{\mu\nu} q_L)\,, \\
Q_1^{LR}&=&(\bar b_L \gamma_\mu q_L)(\bar b_R \gamma^\mu q_R)\,, \\
Q_2^{LR}&=&(\bar b_R q_L)(\bar b_L q_R)\,,
\label{eq:operatorsDF2}
\eea
where $q=s,d$ and $\sigma_{\mu\nu}=\frac{1}{2}\left[\gamma_\mu,\gamma_\nu\right]$. In addition,
the operators $Q_{1,2}^{SRR}$, analogous to $Q_{1,2}^{SLL}$ with the exchange $q_L\to q_R$,
are also present.

Once the Wilson coefficients of these operators $C_i(\mu_H)$ are calculated at a high energy scale,
where heavy degrees of freedom are integrated out, their values at scales $\mathcal O(\mu_B)$
are obtained by the standard techniques~\cite{Buras:2001ra}. The resulting low energy effective
Hamiltonian then reads
\beq
\cH_{\rm eff}=\sum_{i,a}C_i^a(\mu_B,B)Q_i^a~,
\eeq
where $i=1,2$, and  $a=LR,SLL$. The off-diagonal element in B meson mixing are given by
\beq
\label{eq:M12q}
M_{12}^q = \frac{1}{3}M_{B_q} F_{B_q}^2\sum_{i,a} C_i^{a*}(\mu_H,B_q) P_i^a(B_q)\,,
\eeq
where $P_i^a(B_q)$ collect all RG effects from scales below $\mu_H$ as well as hadronic
matrix elements obtained by lattice methods. Updating the results of Ref.~\cite{Buras:2001ra},
it turns out that $P_2^{LR} (B_q)\approx 3.4$ and $P_1^{SLL} (B_q)\approx -1.4$ for
$\mu_H=246$~GeV.

Introducing the notation
\begin{equation}
M_{12}^q=\left(M_{12}^q\right)_{\text{SM}} C_{B_q}e^{2 i\varphi_{B_q}}~,
\qquad (q=d,s)~,
\label{eq:M12}
\end{equation}
the $B_{s,d}$ mass differences and the CP asymmetries $S_{\psi K_S}$ and $S_{\psi\phi}$ 
are
\bea
\Delta M_q &=& 2\left|M_{12}^q\right| = (\Delta M_q)_{\text{SM}}C_{B_q}~,
\label{eq:Delta_Mq} \\
S_{\psi K_S} &=& \sin( 2\beta + 2\varphi_{B_d} )~,\\
S_{\psi\phi} &=& \sin( 2|\beta_s| - 2\varphi_{B_s} )~,
\label{eq:CPV_Bq}
\eea
where $\sin(2\beta)_{\rm tree}=0.734\pm 0.038$~\cite{Bona:2007vi} and
$\sin(2\beta_s)=0.038\pm 0.003$~\cite{Bona:2007vi}.


Moreover, we recall that the semileptonic asymmetry $A^{s}_{\text{SL}}$, in the presence
of NP and neglecting $\beta_s$, is correlated model-independently with $S_{\psi\phi}$
as~\cite{Ligeti:2006pm}
\begin{equation}
\label{eq:A_SL_corr}
A^{s}_{\text{SL}} = -\left |{\rm Re} \left( \frac{\Gamma^{s}_{12}}{M^{s}_{12}}\right)^{\rm SM} \right|\frac{1}{C_{B_s}}S_{\psi\phi}~,
\end{equation}
(for an alternative model-independent formula, see~\cite{Grossman:2009mn})
where $|{\rm Re}(\Gamma^{s}_{12}/M^{s}_{12})^{\rm SM}|=(2.6 \pm 1.0)\times 10^{-3}$~\cite{Ciuchini:2003ww}.


Assuming the $a_i$ in Eq.~(\ref{eq:LH_FCNC}) as complex parameters and integrating out
the neutral Higgs fields we have for the Wilson coefficients of the dominant operators
\bea
\label{eq:M12sHiggsMFV}
C^{SLL}_{1}
\!&\simeq&\!
-\frac{[V_{tb}V^*_{tq}]^2}{4}(a_0+a_1)^2 y^2_t y^2_b~\mathcal{F}_{-}~,
\nonumber\\
C^{LR}_{2}
\!&\simeq&\!
-\frac{[V_{tb}V^*_{tq}]^2}{2}(a_0+a_1)(a^*_0+a^*_2) y^2_t y_b y_q~\mathcal{F}_{+}~, \qquad 
\eea
where, in the decoupling regime ($v^{2}/m_{A}^{2}\ll 1$), and for large $\tan\beta$ values,
\beq
\mathcal{F}^{+}
\simeq
\frac{2}{M_{A}^{2}}\,,\qquad
\mathcal{F}^{-}
\simeq
-\lambda_{5}\frac{v^{2}}{M_{A}^{4}}\,,
\label{F_pm}
\eeq
in agreement with~\cite{Gorbahn:2009pp}.
As can be seen, the $A$--$H$ mixing, related to the  $SU(2)_L$ breaking in the heavy Higgs
sector, removes the zero in $\mathcal{F}^{-}$ providing at the same time a possible source of
CP violation independent from the $a_i$. The mass splitting between $A$ and $H$ is constrained
by the requirement of perturbative unitarity~\cite{Lee:1977yc,Kanemura:1993hm,unitarity2},
vacuum stability~\cite{vacuum-stability} and by precision electroweak observables 
(the oblique-corrections). However, a $(10-15)\%$ splitting for $M_A$ around 500~GeV fulfills
all these constraints.

A close inspection of Eq.~(\ref{eq:M12sHiggsMFV}) reveals that:
\begin{enumerate}
 \item
the contribution of $C^{LR}_{2}$ to $B_q$ mixing is proportional to $m_b m_q$, hence, only
$B_s$ mixing can be affected in a sizable way by $Q_2^{LR}$ (as in the scenario considered in~\cite{Buras:2010mh});
\item
the effect of $C^{SLL}_{1}$ to $B_{s}$ and $B_{d}$ mixings is the same, implying a common
NP phase $\varphi_{B_d}=\varphi_{B_s}$; as a result, the limited room for NP allowed in
$S_{\psi K_S}$ forbids large effects in $S_{\psi\phi}$ and $A^{s}_{SL}$ coming from this operator~\cite{Buras:2008nn,Kagan:2009bn,Blum:2010mj};
\item
$C^{LR}_{2}$ contributes to CP violation only if $a_1 \neq a_2$, which requires effective
operators with high powers of Yukawa insertions; these operators are naturally suppressed
in explicit (perturbative) models, such as supersymmetric models~\cite{ABP,ABGPS}.
However, this suppression can be removed in a general non-supersymmetric 2HDM with MFV.
\item
$C^{SLL}_{1}$ is complex already at the leading order in the Yukawa insertions;
however, it vanishes in the exact $SU(2)$ limit. While in a generic 2HDM rather
sizable $SU(2)$ breaking effects (say at the $10\%$ level) are generally allowed,
in supersymmetric models such effects are loop-induced and therefore too small
to provide any observable effect~\cite{ABP,ABGPS}.
This has also been reemphasized recently in~\cite{Blum:2010mj}.
\end{enumerate}

In order to understand which are the regions of the parameter space where $S_{\psi\phi}$
obtains sizable (non-standard) values, let us first focus on the effects from $C^{LR}_{2}$,
setting $m_A = 500$~GeV and $t_{\beta}=10$. In this case it turns out that~\cite{Buras:2010mh}
\begin{equation}
S_{\psi\phi}\approx
0.15 \times \text{Im}\left[(a_0 + a_1)(a^\star_0 + a^\star_2)\right] + (S_{\psi\phi})_{\rm SM}\,.
\label{eq:Spsiphi_approx}
\end{equation}
Similarly, the SM prediction for BR$(B_q\to \mu^+\mu^-)$ is modified according
to~\cite{D'Ambrosio:2002ex,Buras:2010mh}
\beq
\frac{{\rm Br}(B_{q}\to\mu^+\mu^-)}{{\rm Br}(B_{q}\to\mu^+\mu^-)_{\rm SM}}\simeq
| 1 + R ~|^2 + | R |^2~,
\label{eq:bsmm}
\eeq
where
\beq
R \approx
(a_0^*+a^*_1)\left(\frac{t_\beta}{10}\right)^2
\left(\frac{500\,{\rm GeV}}{M_H}\right)^2~,
\label{eq:bsmm2}
\eeq
with ${\rm Br}(B_d\to\mu^+\mu^-)_{\rm SM} = (1.0\pm 0.1)\times 10^{-10}$ and
${\rm Br}(B_s\to\mu^+\mu^-)_{\rm SM} = (3.2\pm 0.2)\times 10^{-9}$.
Therefore, $S_{\psi\phi}$ can be large, even for moderate values of $t_{\beta}$ and relatively
heavy Higgs masses, provided order one NP phases and sizable PQ-symmetry breaking sources,
i.e.~if $\text{Im}[(a_0 + a_1)(a^\star_0 + a^\star_2)]\approx 1$ are present.
These conditions generally imply large NP effects for BR$(B_q\to\mu^+\mu^-)$, as already
observed in~\cite{Buras:2010mh} and as clearly shown by Eqs.~(\ref{eq:bsmm}),~(\ref{eq:bsmm2}).
As we will show later, also for the EDMs of physical systems like the Thallium (Tl),
Mercury (Hg) and the neutron EDMs large NP effects are expected. In particular, 
in the ${\rm 2HDM_{\overline{MFV}}}$, large non-standard values for $S_{\psi\phi}$ imply
lower bounds for the above EDMs in the reach of the expected future experimental resolutions.

As far as NP effects induced by $C^{SLL}_{1}$ are concerned, we notice that it is
easy to generate a large common phase $\varphi_{B_d}=\varphi_{B_s}$ provided the
mass splitting between $A$ and $H$ is around the $10\%$ level (or above). In this
case the limits on NP effects in $S_{\psi K_S}$ set the bound $S_{\psi\phi}\lesssim 0.2$~\cite{Buras:2008nn,Kagan:2009bn,Blum:2010mj}.
As shown in Eq.~(\ref{eq:M12sHiggsMFV}), $C^{SLL}_{1}$ dominates over $C^{LR}_{2}$ if
we allow only flavour-blind phases in the Higgs potential (${\rm Im}\lambda_5\not=0$)
and we allow a sufficient $SU(2)_L$ breaking in the heavy Higgs sector.


Finally, let us mention that there are also charged Higgs contributions to $M_{12}^q$
that are sensitive to new flavour blind CP sources. These effects have recently been
analyzed in~\cite{Jung:2010ik}, in the context of the 2HDM proposed in~\cite{Pich1}.
They arise from the one loop exchange of $H^{+}-H^{+}$ and $H^{+}-W^{+}$ and lead to
\bea
(C_{1}^{SLL})_{H^+} &=&
\frac{G^2_F}{4\pi^2} m_b^2
\bigg[(C^{H^+}_{L}C^{H^{+*}}_{R})^2 D_0(x_{tH},x_{tH},1) +
\nonumber\\
&-&
2 C^{H^+}_{L}C^{H^{+*}}_{R} D_0(x_{tH},x_{tH},x_{WH})\bigg]~,
\eea
where $D_{0}$ is the standard four-point functions normalized to $D_0(1,1,1)=-1/6$.
In the framework we are considering, with large/moderate $t_\beta$ and small/moderate
$U(1)_{\rm PQ}$ breaking, $(C_{1}^{SLL})_{H^+}$ is always very suppressed and does
not provide visible effects in physical observables. However, this is not the case
in the framework analyzed in Ref~\cite{Jung:2010ik}, where the charged-Higgs contribution
to $B_{s,d}$--$B_{s,d}$ mixing can be sizable in specific regions of the parameter
space (corresponding to large violations of the $U(1)_{\rm PQ}$ symmetry).

\section{5.~~Numerical analysis}

Having introduced the two sets of observables, namely EDMs and $\Delta F=2$ CPV asymmetries,
we are ready to analyse their correlations in the ${\rm 2HDM_{\overline{MFV}}}$.
As outlined in the introduction, we analyse separately the cases of flavour-blind
phases from i) the Yukawa interactions and ii) the Higgs potential.

Starting with case i), in the upper plot of Fig.~\ref{edm_1} we report the expectations
for the EDMs of the Thallium $d_{Tl}$ (red dots), neutron $d_n$ (black dots), and Mercury
$d_{Hg}$ (green dots) as a function of the phase in the $B_s$ mixing, described by the
asymmetry $S_{\psi\phi}$. The plots have been obtained by means of the following scan: $(|\overline{a}|,|a_0|,|a_1|,|a_2|)< 2$, $0<(\phi_{\overline{a}},\phi_{a_0},\phi_{a_1},
\phi_{a_2})< 2\pi$, $\tan\beta < 60$, $\lambda_5=0$, $M_{H^{\pm}}< 1.5$~TeV
and setting the hadronic parameter $\kappa$~\cite{KBM} entering the CP-odd four fermion
coefficients of Eqs.~(\ref{eq:CS}),(\ref{eq:mercuryedm}) to its central value $\kappa = 0.5$.
Moreover, we have imposed the further condition $|a_2|<(|\overline{a}|,|a_0|,|a_1|)$, since
$a_2$ is generated only beyond the leading order in the MFV expansion in terms of the spurions
$Y_u Y_u^\dagger$ and $Y_d Y_d^\dagger$, in contrast to $\overline{a},a_0,a_1$.

As can be seen, the current constraints from the EDMs still allow values of $|S_{\psi\phi}|$
larger than 0.5, compatible with the highest values of the $B_s$ mixing phase reported by
the Tevatron experiments. Yet, sizable non standard values for $S_{\psi\phi}$ unambiguously
imply lower bounds for the above EDMs within the reach of the expected future experimental
resolutions. Similarly, large values for $S_{\psi\phi}$ typically imply a BR$(B_s\to \mu^+\mu^-)$
departing from the SM prediction, as already observed in~\cite{Buras:2010mh} and as clearly
shown in the lower plot in Fig.~\ref{edm_1}.

Moreover, it turns out that BR$(B_d\to \mu^+\mu^-)$/BR$(B_s\to \mu^+\mu^-)\approx
|V_{td}/V_{ts}|^2$, as expected by the model-independent analysis within the MFV
framework of Ref.~\cite{Hurth:2008jc}.

From these plots we conclude that improvements of the experimental lower bounds for the
Thallium, neutron and Mercury EDMs, as well as for BR$(B_{s,d}\to \mu^+\mu^-)$, together
with a more accurate measurement of $S_{\psi\phi}$, would provide a powerful tool in the
attempt to test or to falsify this NP model.

In Fig.~\ref{edm_2}, we show the predictions for $S_{\psi K_S}$ vs. $S_{\psi\phi}$
where the allowed ranges for NP effects in $S_{\psi K_S}$ have been obtained combining
the SM prediction $\sin(2\beta)_{\rm tree}=0.734\pm 0.038$~\cite{Bona:2007vi} with the
experimental result $S^{\rm exp}_{\psi K_S}=0.672\pm 0.023$~\cite{Barberio:2008fa}.
In the upper plot, we switch on the CPV phases of the Yukawa couplings and the Higgs
potential separately, to monitor their individual effect.
In the former case, we employ the same scan as in Fig.~\ref{edm_1}, while in the latter
case we make the scan $(|\overline{a}|,|a_0|,|a_1|,|a_2|)< 2$,
$(\phi_{\overline{a}},\phi_{a_0},\phi_{a_1},\phi_{a_2})=0$, $|\lambda_5|=0.1$
and $0<\phi_{\lambda_5}<2\pi$.
Viceversa, in the lower plot  of Fig.~\ref{edm_2}, we consider an {\it hybrid} scenario,
where CPV phases of both type i) and type ii) are switched on simultaneously.
In both plots, red dots fulfill the EDM constraints while the black ones do not.

The correlation of the EDMs vs.~$S_{\psi\phi}$ in the case~ii), where the flavour blind CPV
phases originates only from the Higgs potential, does not show appreciable differences with
respect to Fig.~\ref{edm_1}. However, since in this cases the the effects on $B_{s,d}$ mixing
are universal, here the larger values of $S_{\psi\phi}$ are not allowed by the constraints
from $B_d$ mixing. This is clearly illustrated in the upper plot of Fig.~\ref{edm_2}.

Finally, Fig.~\ref{edm_3} shows the correlation between $A^{s}_{SL}/A^{s}_{SL}({\rm SM})$
vs. $S_{\psi\phi}$ for the {\it hybrid} scenario, performing the same scan of Fig.~\ref{edm_2}.
Red dots fulfill the EDM constraints while the black ones do not.
Green dots further satisfy the constraint from $S_{\psi K_S}$ at the $95\%$ CL..
As can be seen, in this scenario sizable/large effects for $B_d/B_s$ mixings can be
naturally accounted for, even though not necessarily in a correlated manner. According
to the recent model-independent fit of the two mixing phases performed in~\cite{Jurenow},
this hybrid scenario is particularly welcome by present data. Indeed the best fit
of current data is obtained for a $\varphi_{B_d}/\varphi_{B_s}$ ratio which is in
between the $\varphi_{B_d}\approx(m_d/m_s) \varphi_{B_s}$ of the scenario~i) and
the $\varphi_{B_d} \approx \varphi_{B_s}$ of the scenario~ii).

Finally, as demonstrated in Fig.3 of~\cite{Buras:2010mh} for large values of $S_{\psi\phi}$
a unique positive shift in $\varepsilon_K$ is implied within the model considered, bringing
the theory closer to the data. This agreement is further improved through the recently
calculated NNLO QCD corrections to $\varepsilon_K$~\cite{Brod:2010mj}. In the hybrid
scenario, a value for  $\varepsilon_K$ within 1~$\sigma$ of current data can be obtained
even for values of $S_{\psi\phi}\approx 0.25$.

\section{6.~~Discussion and conclusions}

The remarkable agreement of flavour data with the SM predictions in the $K$ and $B_d$
systems highly constrains the flavour structure of any TeV-scale UV completion of the SM.
In this respect the MFV hypothesis~\cite{D'Ambrosio:2002ex}, where flavour-changing
phenomena are entirely controlled by the CKM matrix, represents one of the most natural
and elegant explanations for such an impressive agreement. However, the MFV principle
does not forbid in itself the presence of new flavour blind CP violating phases in addition
to the unique phase of the CKM matrix~\cite{Kagan:2009bn,Mercolli:2009ns,Paradisi:2009ey}.

From a phenomenological point of view, it is natural to expect that such 
flavour-blind phases, if present, will contribute not only to 
flavour-changing CPV processes, such as CP-violation in $B_{s,d}$ mixing,
but also to flavour-conserving CPV processes such as the EDMs.
Indeed the close connection within a MFV framework between CP violation
in flavour physics and EDMs has already been analyzed in the context of
supersymmetric extensions of the SM~\cite{ABP,ABGPS}, and by 
means of a general effective theory approach~\cite{Batell:2010qw}.

In this Letter we have analyzed these connections within the 2HDM respecting the MFV
hypothesis recently discussed in~\cite{Buras:2010mh}. We have shown that the two classes
of flavour-blind phases present in this model, those in the Yukawa couplings and those
in the Higgs potential, provide potentially large CP violating effects in $B_{s,d}$
mixing while being compatible with the EDM bounds of the neutron and heavy atoms.
In both cases sizable CPV effects in $B_{s}$ mixing imply lower bounds for the
above EDMs within the future experimental resolutions. Moreover, in both cases
${\rm BR}(B_{s,d}\to\mu^+\mu^-)$ are typically largely enhanced over their SM
expectations.

What distinguishes the different flavour-blind CPV mechanisms 
is the correlation between  $S_{\psi K_S}$ and, correspondingly,
the maximal effect expected in $S_{\psi\phi}$. Very large values 
of $S_{\psi\phi}$ are possible only via the mechanism pointed out 
in~\cite{Buras:2010mh}, which requires flavour-blind phases in 
the Yukawa interactions and small $SU(2)_L$ breaking 
in the heavy Higgs sector (or the dominance of the effective 
operator $Q_2^{LR}$). On the contrary, flavour-blind phases 
only in the Higgs potential leads to universal effects in 
$B_s$ and $B_d$ mixing, which are strongly constrained by the 
measurement of $S_{\psi K_S}$.
Which of the two flavour-blind CPV mechanisms dominates 
depends on the precise values of $S_{\psi\phi}$ 
and $S_{\psi K_S}$, as well as on the CKM phase (as determined 
by tree-level processes). Current data seems to 
show a mild preference for an  {\it hybrid} scenario
where both these mechanisms are at work.

\section*{Acknowledgments}
We thank M.~Nagai for valuable discussions and for his critical reading of the manuscript.
This research was partially supported by the Cluster of Excellence `Origin and Structure
of the Universe', by the German `Bundesministerium f\"ur Bildung und Forschung'
under contract 05H09WOE, and by the EU Marie Curie Research Training Network contracts
MTRN-CT-2006-035482 ({\em Flavianet}).


\end{document}